\documentclass[english]{article}
\usepackage[LGR,T1]{fontenc}
\usepackage[utf8]{inputenc}
\usepackage{color}
\usepackage{babel}
\usepackage{array}
\usepackage{float}
\usepackage{multirow}
\usepackage{amstext}
\usepackage{graphicx}
\usepackage{geometry}
\usepackage{authblk}
\usepackage{orcidlink}
\usepackage{authblk}
\usepackage{verbatim}
\usepackage{esint}
\usepackage{hyperref}
\usepackage{cite}
\hypersetup{
    colorlinks=true,
    linkcolor=blue,
    filecolor=magenta,      
    urlcolor=cyan,
    pdftitle={Overleaf Example},
    pdfpagemode=FullScreen,
    }

\title{Spatio-Temporal Synchronization of Counter-Propagating Femtosecond Pulses}

\author[1,2,*]{Tamir Cohen\orcidlink{0000-0003-2497-1613}}
\author[2]{Moshe Fraenkel\orcidlink{0000-0002-9193-9544}}
\author[1]{Ishay Pomerantz\orcidlink{0000-0001-9824-887X}}

\affil[1]{The School of Physics and Astronomy, Tel Aviv University, Tel Aviv 69978, Israel}

\affil[2]{Plasma Physics Department, Soreq NRC, Yavne 81800, Israel}

\affil[*]{Author to whom any correspondence should be addressed.}

\providecommand{\keywords}[1]
{
  \small	
  \textbf{Keywords:} #1
}
\providecommand{\email}[1]
{
  \small	
  \textbf{E-mail: } #1
}

\begin{document}
\maketitle

\noindent\email{tamircohen3@mail.tau.ac.il}\\

\noindent\keywords{
   spatio-temporal synchronization, ultrashort laser pulses, counter-propagating beams, wavefront sensing; interferometry, plasma-guided Compton source.}

\begin{abstract}
\noindent Generating intense x-ray radiation via inverse-Compton scattering and exploring the strong-field regime of QED, require precise spatio-temporal synchronization of tightly focused counter-propagating intense laser pulses. We present a protocol for establishing spatio-temporal synchronization in this geometry that combines microscope-based target positioning, wavefront-sensor-assisted alignment of off-axis parabolic mirrors, and a high-resolution temporal delay scan based on interference. We observed an interference window of 72~fs in good agreement with the expected autocorrelation width. The accuracy levels in space and time of using this protocol are discussed.
\end{abstract}

\section{Introduction}

Counter-propagating tightly focused laser beams are the basis for several high-intensity laser--plasma experiments, like strong-field quantum electrodynamics (SFQED) \cite{cole2018experimental}  and laser-driven inverse Compton scattering (ICS)\cite{compton1923quantum,Alejo2019Laser,Petrillo2023State}. For SFQED, the counter-propagating pulses form a high-field interaction region in which nonlinear quantum-electrodynamic processes, including nonlinear Compton scattering \cite{bamber1999studies, mirzaie2024all}, radiation reaction\cite{gonoskov2022charged, wistisen2019quantum,poder2018experimental}, and electron--positron pair production \cite{xie2017electron}, can be investigated  . In ICS, a laser accelerated  electron beam collides head-on with a counter-propagating ultrashort laser pulse, upshifting the laser photons to the x-ray or gamma-ray spectral range \cite{chen2013mev, albert2023principles}. Both configurations therefore require precise transverse overlap and temporal synchronization of the focused beams at the interaction point. For counter-propagating, tightly focused pulses, this requires spatial alignment to within a few microns and temporal synchronization on the femtosecond scale, since small  deviations can substantially reduce the interaction strength at the interaction point.

\begin{figure}[htbp]
    \centering
    \includegraphics[width=14. cm]{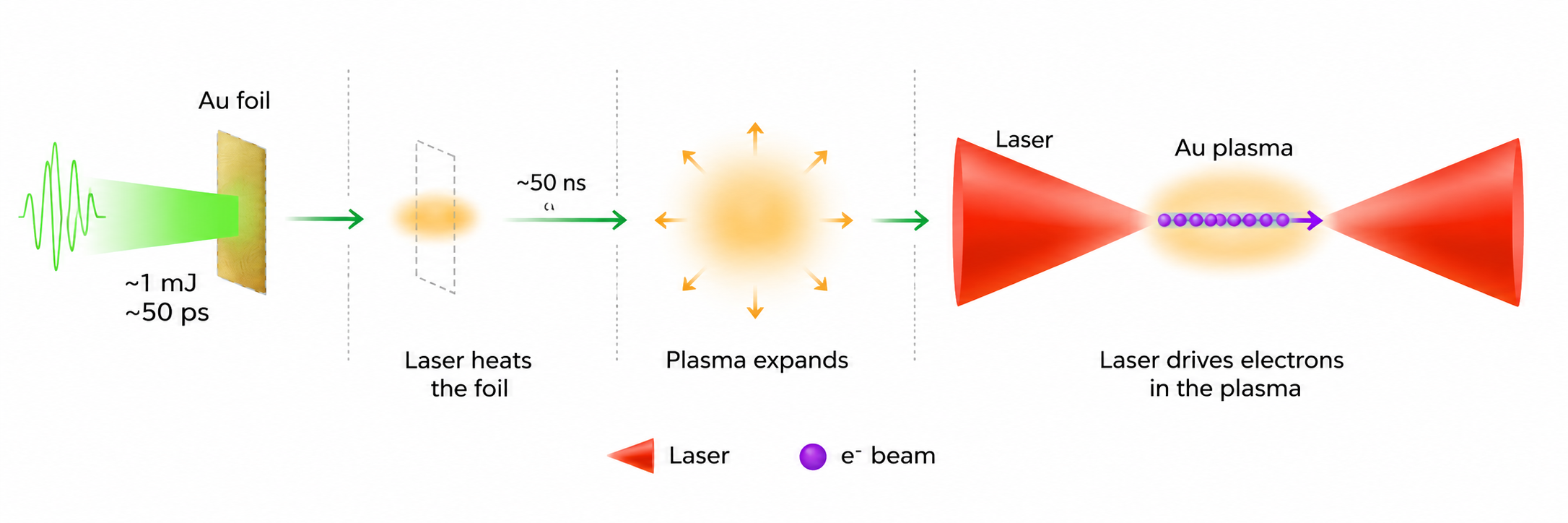}
    \caption{Cartoon schematic of the PGCS scheme. A plasma plume is formed by irradiating a thin Au foil with a 50~ps laser pulse delivering $\sim$1~mJ of energy. After a 50-ns expansion interval, the plume is irradiated by two counter-propagating laser pulses at relativistic intensity.}
    \label{fig:PGCS schem}
\end{figure}

In our recent numerical work  we proposed the "plasma-guided Compton source (PGCS)" \cite{meir2024plasma} ICS scheme, illustrated in Fig.~\ref{fig:PGCS schem}.
It is based on the direct laser acceleration (DLA) \cite{pukhov1999particle, arefiev2016beyond,cohen2024undepleted,cohen2026stabilizing,cohen2024multi} method, in which electrons are accelerated to MeV energies in a near-critical-density plasma.

Temporal synchronization of ultrashort laser pulses is well established using non‑linear optical cross‑correlation in BBO and related crystals, operating in either collinear or non‑collinear geometries, where the relative delay between two ultrashort pulses is retrieved from the generated sum‑frequency (SFG) or second‑harmonic signal (SHG) \cite{liu2021timing,schibli2003attosecond}. 
However, such techniques are unsuited for counter-propagating laser beams, for which the opposing propagation directions are incompatible with conventional phase-matching geometries.
An alternative optical technique based on spectrally resolved interferometry with a pellicle and diffraction grating has been demonstrated to achieve femtosecond‑scale synchronization of ultra‑intense focused laser beams \cite{corvan2014femtosecond}. However, this technique is limited to transparent substrates and is not applicable to plasma experiments using the exploding-foil method of opaque materials.

In this work, we describe a proof‑of‑concept alignment protocol that achieves spatio-temporal synchronization between two counter‑propagating ultrashort pulses in a tightly focused geometry directly relevant to PGCS experiments.  The protocol combines microscope-based target positioning, wavefront-sensor-assisted beam alignment, and a temporal scan to establish the interaction point with micrometric precision and synchronize the pulses at that point.

\section{Experimental setup}

This work was conducted using the 30 TW OPCPA \cite{cohen2026stabilizing} laser system at the Soreq Nuclear Research Center (SNRC). The laser delivers pulses centered at $\lambda\approx 840$~nm with tunable pulse durations ranging from 35~fs to 1000~fs, and up to 1~J of energy at a repetition rate of 5~Hz. For the alignment measurements reported here, the pulse energy was substantially attenuated. The pulses were first stretched to 1000~fs to facilitate coarse delay-line scans, then recompressed to 35~fs for fine synchronization measurements.

\begin{figure} [htbp]
    \centering
    \includegraphics[width=0.75\linewidth]{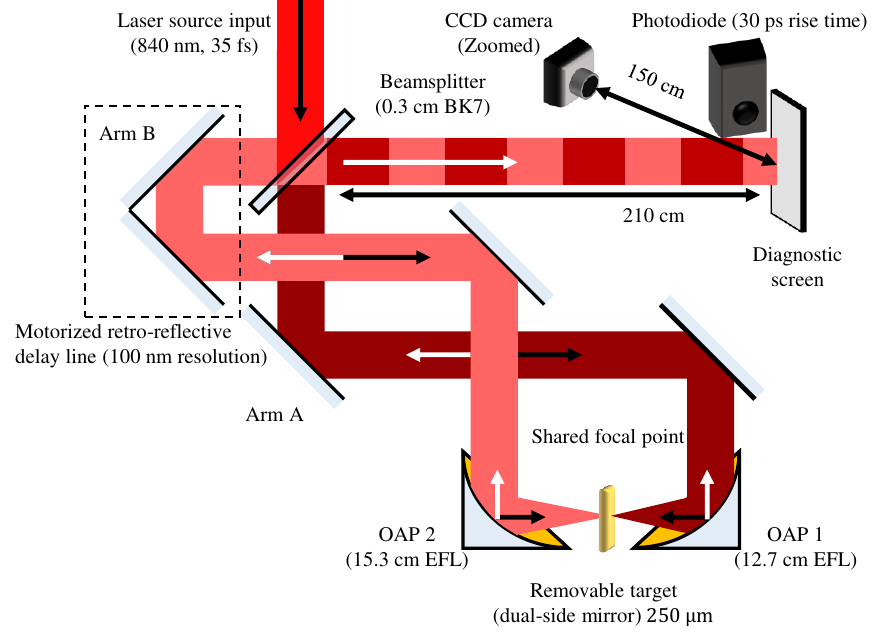}
    \caption{A cartoon schematic of the experimental setup. The incoming laser beam is split into Arms A and B, routed through a delay line in Arm B, and focused by two off-axis parabolic mirrors onto a common target (act as dual side mirror) at the shared focal point. The reflected beams are then recombined and monitored on a downstream diagnostic screen using a fast photodiode and a CCD camera during temporal alignment scans. }
    \label{fig:System schem}
\end{figure}

Fig. \ref{fig:System schem} shows a schematic of the experimental setup. An incoming beam is split into Arm A and Arm B by a 3~mm BK7 beamsplitter at an incidence angle of 45$^\circ$. Arms A and B are routed to two 90$^\circ$ off-axis parabolic (OAP) mirrors with effective focal lengths (EFL) of 12.7~cm and 15.3~cm, respectively. The OAP mirrors are mounted to 6-axis kinematic stages. A retro-reflecting optical delay line, mounted on a linear stage with a spatial resolution of 100~nm, is used to vary the optical path length in Arm B. The two OAP mirrors are aligned so that they share a common focal spot, at which a dual‑sided mirror (target) is placed to reflect both counter‑propagating beams.

The beams reflected from the target retraced their paths towards the beamsplitter, recombined, and interfered on a diagnostic screen located 210~cm downstream of the beamsplitter when the two beams are synchronized. Temporal overlap is first established through a coarse delay scan monitored with a fast InGaAs photodiode. A finer delay scan is then performed while interference fringes are recorded with a CCD camera. The appearance of these fringes provides a sensitive indication of spatio-temporal overlap.

\section{Spatio-temporal alignment protocol}
\begin{figure}[htbp]
    \centering
    \includegraphics[width=0.92\linewidth]{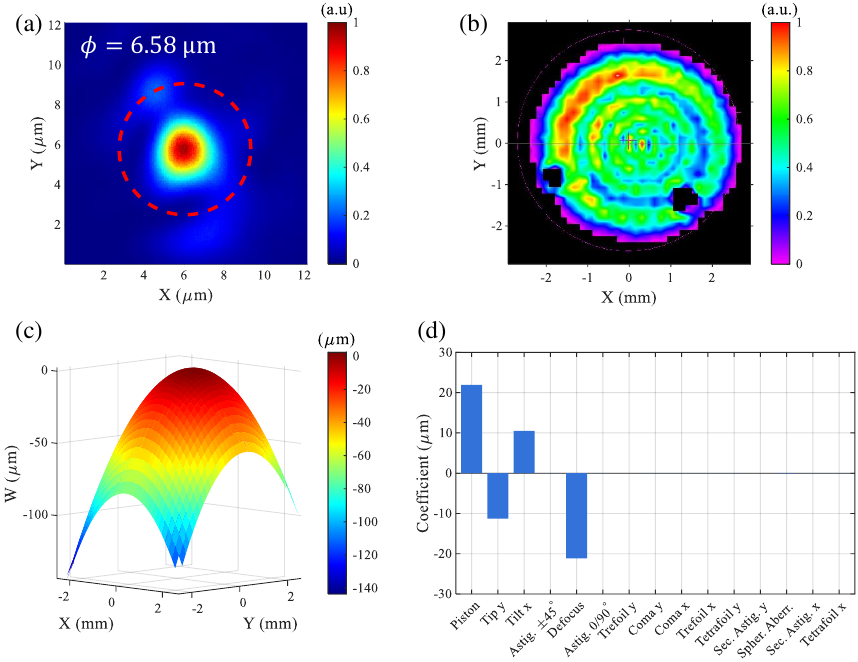}
    \caption{(a) Focal spot recorded by the $40\times$ magnifying microscope. The red dashed contour encloses the region containing 70$\%$ of the energy recorded in the focal-spot image. (b) WFS measurement acquired out of focus. (c) Measured wavefront of the focused beam, demonstrating high defocus. (d) Extracted Zernike coefficients used to quantify the wavefront aberrations. Aberrations of high order demonstrated values $<|0.09|$~\textmu{m} }
    \label{fig:Focal Spot}
\end{figure}

In Step-1 of the alignment procedure, the coordinates of the interaction point are defined. An un-split beam is injected along Arm A into OAP-1 (EFL~$=$~12.7~cm), focused, and inspected using a ×40 microscope while the OAP mirror is adjusted until a symmetric focal spot with minimal aberrations is obtained (Fig. \ref{fig:Focal Spot}.a). The focal region is also characterized using a Shack–Hartmann wavefront sensor to verify wavefront quality (Fig. \ref{fig:Focal Spot}.b–d).  

The beamsplitter is then inserted, and the focal spot is readjusted if necessary. Once the focal point is optimized, the microscope is left in place and used as a spatial reference for positioning the reflective target (double-sided mirror). The target is mounted to an encoded 6-axis stage and translated until its surface is brought into focus. For this purpose, the microscope is illuminated by a continuous-wave (CW) 840~nm diode laser injected from the back of the objective, so that it focuses at the same position as the OAP focal spot. This procedure places the target surface in the OAP focal plane, as the depth of field (DoF) of the microscope (1.6~µm) is much smaller than the Rayleigh range of the OAPs ($>$38~µm).

\begin{figure}[htbp]
    \centering
    \includegraphics[width=0.9
    \linewidth]{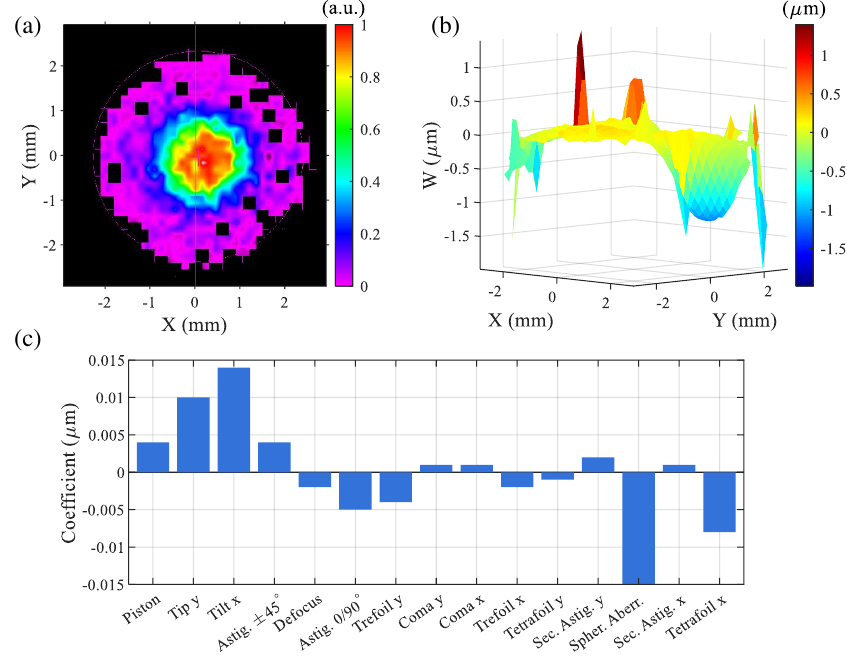}
    \caption{ Characterization of outgoing beam from OAP-2 using  a Shack–Hartmann wavefront sensor. (a) Beam intensity profile(b) Measured wavefront, showing minimal deviation from a flat wavefront. (c) Extracted Zernike coefficients used to quantify the residual wavefront aberrations.}
    \label{fig:Colimated beam}
\end{figure}

In Step-2, the opposing beam path is established and aligned. The target is removed, and the alignment is performed using the beam injected from Arm A. This beam is directed onto OAP-2 (EFL~$=$~15.3~cm), whose position and angle are adjusted while the outgoing beam is monitored with a Shack–Hartmann wavefront sensor. The OAP and upstream folding mirror are iteratively tuned until the measured wavefront is nearly flat and the Zernike coefficients are minimized (Fig. \ref{fig:Colimated beam}). This procedure ensures that the beam delivered from OAP-2 is well collimated and free of astigmatism. 

\begin{figure}[htbp]
    \centering
    \includegraphics[width=0.35\linewidth]{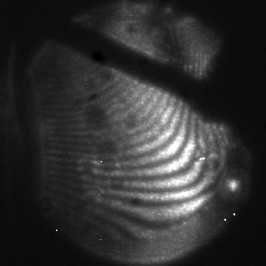}
    \caption{Interference pattern recorded on the diagnostic screen after the counter-propagating beams complete a full round trip through the optical system. The observed fringes indicate spatial and temporal overlap of the recombined beams.}
    \label{fig:system schem full round and interference pattern}
\end{figure}

The incoming and outgoing beams on Arm B are then directed through the same pair of pinholes that define a fixed optical axis between OAP-2 and the beamsplitter in both propagation directions. This step establishes an equal optical propagation path for the counter‑propagating beams. The equality of the optical paths of the counter-propagating beams is verified on the diagnostic screen. Fig.\ref{fig:system schem full round and interference pattern} shows the interference pattern as recorded on the diagnostic screen. This pattern remains unchanged as the delay line is scanned, indicating that both beams traverse identical optical paths within the common‑path section.

In Step-3 an interferometer is formed. The target (double‑sided mirror) is returned to the position defined in Step-1, and its orientation is adjusted so the reflected beams are spatially overlapped. Because changing the target orientation can shift its position, it is necessary to re‑confirm its new position using the microscope procedure described above. Once the target is repositioned and the two beams overlap on the screen, a delay scan is performed. 

The initial coarse scan is carried out with pulse duration stretched to 1000~fs using a fast photodiode with a 30~ps rise time. At large delays of tens of picoseconds, two distinct peaks corresponding to the two pulses are observed on the oscilloscope. As the time delay between the two pulses is reduced, the peaks merge into one, and further delay changes are no longer distinguishable. The delay-line resolution is then increased, and a finer scan is performed while the interference pattern is monitored with a CCD camera to identify the temporal overlap window.

\section{Results}

In the following, we verify the performance of our spatio-temporal synchronization protocol by imaging the interference pattern formed when the two pulses are reflected from a thin target (dual side mirror), while the optical delay is scanned. 

\begin{figure}[htbp]
    \centering
    \includegraphics[width=11.5 cm]{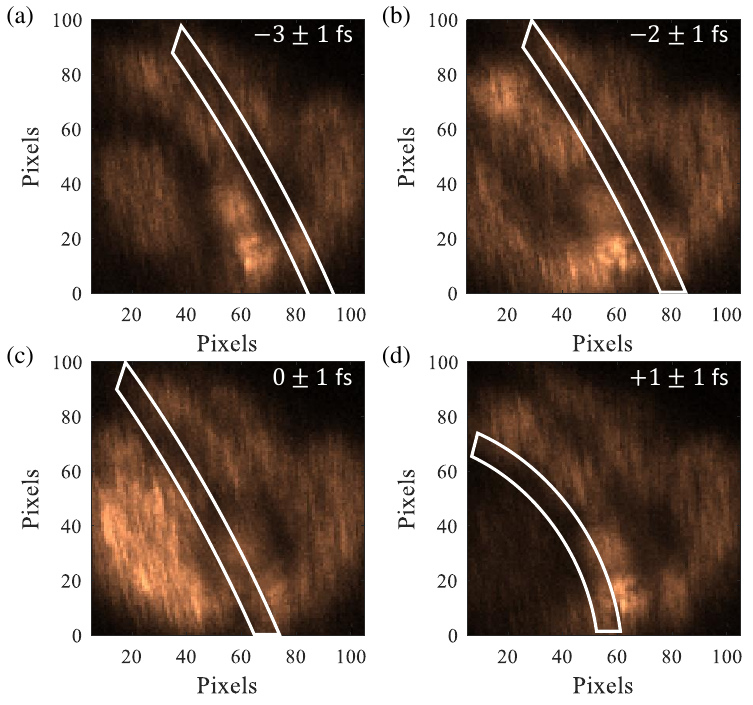}
    \caption { Interference patterns displayed at delays in the range of -3~--~+1~fs. The regions marked in white indicate the same fringe minima and
illustrates its evolution as the relative delay is varied. 
    The $\pm1$~fs uncertainty reflects the temporal sampling associated with the stage motion and the finite laser repetition rate.
    }
    \label{fig:Spatio-temporal alignment}
\end{figure}

Fig. \ref{fig:Spatio-temporal alignment} shows the evolution of the interference pattern in the vicinity of full temporal overlap for four corrected delays in the range of (-3~--~1)$\pm1$~fs. In each panel, a white region highlights the same fringe minimum, providing a visual reference for tracking its evolution as the relative delay is varied.

The measured fringe pattern demonstrates that the synchronization protocol performs as anticipated. The fringes appear only within a temporal window of $72 \pm 1$~fs and disappear outside this window. Using the measured spectrum, the expected coherence length is evaluated as $L_{ac}\approx \lambda^2/\Delta\lambda$, yielding approximately 20~\textmu{m}, translating to an expected $1/e^2$ temporal autocorrelation of 75~fs. The experimentally observed 72~fs interference window is in excellent quantitative agreement with this theoretical value, confirming the high precision and reliability of the spatio-temporal alignment protocol.

\section{Analytical corrections}

The measured interference fringes establish spatio-temporal synchronization within the interferometer geometry. However, PGCS experiments require synchronization of the pulse arrival times at the physical interaction point inside the plasma plume ablated from the target. The optical paths in these two configurations are not identical. In the interferometer configuration, each beam undergoes one transmission and one reflection at the beamsplitter. At the interaction point, however, the beam in Arm A is only transmitted whereas the beam in Arm B is only reflected. Consequently, the two arms acquire different group delays. To account for this, we evaluate the additional group delay introduced by the beamsplitter in Arm A, in terms of optical path \cite{hecht2002optics}.
\begin{equation}
    \Delta \text{OPL}_{group} = d\left(\sqrt{n^2_g - \sin^2(\theta_i)} - \cos(\theta_i) \right) \approx 1.91\text{ mm}
\end{equation}
where $d=3\text{~mm}$ is the thickness of the beamsplitter, $n_g = 1.52$ is the group refractive index of the beamsplitter's material (BK7), and $\theta_i = 45^\circ$ is the incident angle. This corresponds to temporal difference of  $\Delta t = \Delta \text{OPL}_{group}/c = 6.37\text{~ps}$.

A second correction than need to be considered accounts for the target thickness. During diagnostic alignment, the beam in Arm A is reflected from the front surface of the target, which is \(250~\mu\mathrm{m}\) upstream of the interaction plane defined at the rear surface. Transitioning to the physical experiment requires the beam in Arm A to propagate this additional $250~\mu\text{m}$ to the interaction point, introducing an extra temporal delay of $0.83\text{ ps}$. Because both the group delay ($6.37\text{ ps}$) and target offset ($0.83\text{ ps}$) delay Arm A, achieving true spatio-temporal synchronization requires a total cumulative correction of $7.20\text{ ps}$. To compensate for this offset, the optical path in Arm B is increased by $2.16\text{ mm}$, physically implemented via a $1.08\text{ mm}$ translation of its motorized retro-reflector delay stage ($\Delta x_{\text{stage}} = \Delta\text{OPL}/2$).

\section{Conclusions}

This work represents an important milestone toward a physical realization of the PGCS experiment. It establishes a complete and reproducible alignment protocol for the future realization of this experiment. The protocol that combines a microscope to define the interaction point, a Shack--Hartmann wavefront sensor for OAP alignment, a two‑step temporal scan, and analytical correction of the differential optical path between the diagnostic and physical interaction geometries. Together, these steps enable micrometer-scale positioning of the interaction point and spatio-temporal synchronization of counter-propagating femtosecond pulses at that point, thereby establishing the conditions required for future PGCS experiments.

In future work, the thick glass beamsplitter will be replaced by a micrometer-scale pellicle. This modification is expected to reduce the beamsplitter-induced differential group delay from the present value to a few femtoseconds. A thinner beamsplitter will also reduce dispersion and nonlinear effects associated with the propagation of intense ultrashort pulses through bulk glass. The target-thickness correction will remain geometry-dependent and will need to be accounted for separately.

\section*{Funding}
This research was supported by Grant No. 2022322
from the United States-Israel Binational Science Foundation (BSF).

\section*{Author contributions}
\noindent Tamir Cohen\orcidlink{0000-0003-2497-1613}: Conceptualization (lead), Methodology (lead), Investigation (lead), Formal analysis (lead), Visualization (lead), Writing -- original draft (lead), Writing -- review \& editing (equal). \\[1\baselineskip]
\noindent Moshe Fraenkel\orcidlink{0000-0002-9193-9544}: Supervision (equal), Resources (equal).\\[1\baselineskip]
\noindent Ishay Pomerantz\orcidlink{0000-0001-9824-887X}: Methodology (supporting), Supervision (equal), Writing -- review \& editing (equal).

\section*{Conflict of Interest}
The authors declare that they have no conflict of interest.

\bibliographystyle{unsrt}
\bibliography{Bibliography}

@article{cole2018experimental,
  title={Experimental evidence of radiation reaction in the collision of a high-intensity laser pulse with a laser-wakefield accelerated electron beam},
  author={Cole, JM and Behm, KT and Gerstmayr, E and Blackburn, TG and Wood, JC and Baird, CD and Duff, Matthew J and Harvey, Christopher and Ilderton, Antony and Joglekar, AS and others},
  journal={Physical Review X},
  volume={8},
  number={1},
  pages={011020},
  year={2018},
  publisher={APS}
}

@article{compton1923quantum,
  author    = {Compton, Arthur H.},
  title     = {A Quantum Theory of the Scattering of X-Rays by Light Elements},
  journal   = {Physical Review},
  volume    = {21},
  number    = {5},
  pages     = {483--502},
  year      = {1923},
  month     = {May},
  publisher = {American Physical Society},
  doi       = {10.1103/PhysRev.21.483}
}

@article{alejo2019laser,
  title={Laser-wakefield electron beams as drivers of high-quality positron beams and inverse-Compton-scattered photon beams},
  author={Alejo, Aaron and Samarin, Guillermo M and Warwick, Jonathan R and Sarri, Gianluca},
  journal={Frontiers in Physics},
  volume={7},
  pages={49},
  year={2019},
  publisher={Frontiers Media SA}
}

@article{petrillo2023state,
  title={State of the art of high-flux Compton/Thomson x-rays sources},
  author={Petrillo, Vittoria and Drebot, Illya and Ruijter, Marcel and Samsam, Sanae and Bacci, Alberto and Curatolo, Camilla and Opromolla, Michele and Conti, Marcello Rossetti and Rossi, Andrea Renato and Serafini, Luca},
  journal={Applied Sciences},
  volume={13},
  number={2},
  pages={752},
  year={2023},
  publisher={MDPI}
}

@article{bamber1999studies,
  title={Studies of nonlinear QED in collisions of 46.6 GeV electrons with intense laser pulses},
  author={Bamber, C and Boege, SJ and Koffas, T and Kotseroglou, T and Melissinos, AC and Meyerhofer, DD and Reis, DA and Ragg, W and Bula, C and McDonald, KT and others},
  journal={Physical Review D},
  volume={60},
  number={9},
  pages={092004},
  year={1999},
  publisher={APS}
}

@article{mirzaie2024all,
  title={All-optical nonlinear Compton scattering performed with a multi-petawatt laser},
  author={Mirzaie, Mohammad and Hojbota, Calin Ioan and Kim, Do Yeon and Pathak, Vishwa Bandhu and Pak, Tae Gyu and Kim, Chul Min and Lee, Hwang Woon and Yoon, Jin Woo and Lee, Seong Ku and Rhee, Yong Joo and others},
  journal={Nature Photonics},
  volume={18},
  number={11},
  pages={1212--1217},
  year={2024},
  publisher={Nature Publishing Group UK London}
}

@article{gonoskov2022charged,
  title={Charged particle motion and radiation in strong electromagnetic fields},
  author={Gonoskov, A and Blackburn, TG and Marklund, M and Bulanov, SS},
  journal={Reviews of Modern Physics},
  volume={94},
  number={4},
  pages={045001},
  year={2022},
  publisher={APS}
}

@article{wistisen2019quantum,
  title={Quantum radiation reaction in aligned crystals beyond the local constant field approximation},
  author={Wistisen, TN and Di Piazza, A and Nielsen, CF and S{\o}rensen, AH and Uggerh{\o}j, UI and CERN NA63},
  journal={Physical Review Research},
  volume={1},
  number={3},
  pages={033014},
  year={2019},
  publisher={APS}
}

@article{poder2018experimental,
  title={Experimental signatures of the quantum nature of radiation reaction in the field of an ultraintense laser},
  author={Poder, Kristjan and Tamburini, Matteo and Sarri, G and Di Piazza, Antonino and Kuschel, S and Baird, CD and Behm, K and Bohlen, S and Cole, JM and Corvan, DJ and others},
  journal={Physical Review X},
  volume={8},
  number={3},
  pages={031004},
  year={2018},
  publisher={APS}
}

@article{xie2017electron,
  title={Electron-positron pair production in ultrastrong laser fields},
  author={Xie, Bai Song and Li, Zi Liang and Tang, Suo},
  journal={Matter and Radiation at Extremes},
  volume={2},
  number={5},
  pages={225--242},
  year={2017},
  publisher={AIP Publishing}
}

@article{chen2013mev,
  title={MeV-energy X rays from inverse Compton scattering with laser-wakefield accelerated electrons},
  author={Chen, S and Powers, ND and Ghebregziabher, I and Maharjan, CM and Liu, C and Golovin, G and Banerjee, S and Zhang, J and Cunningham, N and Moorti, A and others},
  journal={Physical review letters},
  volume={110},
  number={15},
  pages={155003},
  year={2013},
  publisher={APS}
}

@article{albert2023principles,
  title={Principles and applications of x-ray light sources driven by laser wakefield acceleration},
  author={Albert, F{\'e}licie},
  journal={Physics of Plasmas},
  volume={30},
  number={5},
  year={2023},
  publisher={AIP Publishing}
}

@article{meir2024plasma,
  title={Plasma-guided Compton source},
  author={Meir, Talia and Cohen, Itamar and Tangtartharakul, Kavin and Cohen, Tamir and Fraenkel, Moshe and Arefiev, Alexey V and Pomerantz, Ishay},
  journal={Physical Review Applied},
  volume={22},
  number={4},
  pages={044004},
  year={2024},
  publisher={APS}
}

@article{pukhov1999particle,
  title={Particle acceleration in relativistic laser channels},
  author={Pukhov, A and Sheng, Z-M and Meyer-ter-Vehn, J},
  journal={Physics of Plasmas},
  volume={6},
  number={7},
  pages={2847--2854},
  year={1999},
  publisher={American Institute of Physics}
}

@article{arefiev2016beyond,
  title={Beyond the ponderomotive limit: Direct laser acceleration of relativistic electrons in sub-critical plasmas},
  author={Arefiev, AV and Khudik, VN and Robinson, APL and Shvets, G and Willingale, L and Schollmeier, M},
  journal={Physics of Plasmas},
  volume={23},
  number={5},
  year={2016},
  publisher={AIP Publishing}
}

@article{cohen2024undepleted,
  title={Undepleted direct laser acceleration},
  author={Cohen, Itamar and Meir, Talia and Tangtartharakul, Kavin and Perelmutter, Lior and Elkind, Michal and Gershuni, Yonatan and Levanon, Assaf and Arefiev, Alexey V and Pomerantz, Ishay},
  journal={Science advances},
  volume={10},
  number={2},
  pages={eadk1947},
  year={2024},
  publisher={American Association for the Advancement of Science}
}

@article{cohen2026stabilizing,
  title={Stabilizing direct laser acceleration with long-scale-length plasma targets},
  author={Cohen, Tamir and Meir, Talia and Cohen, Itamar and Levanon, Assaf and Grepl, Filip and Tryus, Maksym and Istokskaia, Valeria and Schillaci, Francesco and Giuffrida, Lorenzo and Fraenkel, Moshe and others},
  journal={Applied Physics Letters},
  volume={128},
  number={15},
  year={2026},
  publisher={AIP Publishing}
}

@article{cohen2024multi,
  title={Multi-scale analytical description of an expanding plasma slab},
  author={Cohen, Itamar and Meir, Talia and Elkind, Michal and Catabi, Tomer and Henis, Zohar and Perelmutter, Lior and Pomerantz, Ishay},
  journal={Physics of Plasmas},
  volume={31},
  number={1},
  year={2024},
  publisher={AIP Publishing}
}

@article{liu2021timing,
  title={Timing Fluctuation Correction of A Femtosecond Regenerative Amplifier},
  author={Liu, Keyang and Li, Hongyang and Wang, Xinliang and Liu, Yanqi and Song, Liwei and Leng, Yuxin},
  journal={Crystals},
  volume={11},
  number={10},
  pages={1242},
  year={2021},
  publisher={MDPI}
}

@article{schibli2003attosecond,
  title={Attosecond active synchronization of passively mode-locked lasers by balanced cross correlation},
  author={Schibli, TR and Kim, Jungwon and Kuzucu, O and Gopinath, JT and Tandon, SN and Petrich, GS and Kolodziejski, LA and Fujimoto, JG and Ippen, EP and Kaertner, FX},
  journal={Optics Letters},
  volume={28},
  number={11},
  pages={947--949},
  year={2003},
  publisher={Optical Society of America}
}

@article{corvan2014femtosecond,
  title={Femtosecond-scale synchronisation of ultra-intense focused laser beams},
  author={Corvan, DJ and Schumaker, W and Cole, J and Ahmed, H and Krushelnick, K and Mangles, SPD and Najmudin, Z and Symes, D and Thomas, AGR and Yeung, M and others},
  journal={arXiv preprint arXiv:1409.4243},
  year={2014}
}

@book{hecht2002optics,
  title={Optics, 5e},
  author={Hecht, Eugene},
  year={2002},
  publisher={Pearson Education India}
}

\end{document}